\documentclass[prb,aps,twocolumn,english,notitlepage]{revtex4-1} 
\usepackage[T1]{fontenc} 
\usepackage{sidecap}
\usepackage[latin1]{inputenc}
 \usepackage{babel} \usepackage{txfonts}
 \usepackage{epsfig,epsf,psfrag} 
\usepackage{graphicx} 
\usepackage{tcolorbox}
\usepackage{framed}
\usepackage{booktabs,xcolor}
\usepackage{xcolor,colortbl}
\usepackage{afterpage}
\usepackage[colorlinks=true,linktoc=page,linkcolor=magenta,citecolor=cyan]{hyperref}

\definecolor{Gray}{gray}{0.85}
\definecolor{LightCyan}{rgb}{0.88,1,1}
\definecolor{darkpink}{rgb}{0.91, 0.33, 0.5}
\newcolumntype{a}{>{\columncolor{Gray}}c}
\newcolumntype{b}{>{\columncolor{white}}c}
\def\eps{\epsilon}
\def\la{\langle}
\def\ra{\rangle}
\def\ua{\uparrow}
\def\da{\downarrow}

\newcommand{\ie}{{\it i.e.,}}

\usepackage{subfigure}
 \begin{document} 
\title{Amplification of Cooper pair splitting current in a graphene based Cooper pair beam splitter geometry} 
\author{SK Firoz Islam}
\email{firoz@iopb.res.in}
\author{Arijit Saha}
\email{arijit@iopb.res.in}
\affiliation{Institute of Physics, Sachivalaya Marg, Bhubaneswar-751005, India}
\affiliation{Homi Bhabha National Institute, Training School Complex, Anushakti Nagar, Mumbai 400085, India}

\begin{abstract}
Motivated by the recent experiments [Scientific reports, {\bf 6}, 23051 (2016), Phys. Rev. Lett. 114, 096602 (2015)],
we theoretically investigate Cooper pair splitting current in a graphene based Cooper pair beam splitter
geometry. By considering the graphene based superconductor as an entangler device, instead of normal (2D) BCS superconductor,
we show that the Cooper pair splitting current mediated by Crossed Andreev process is amplified compared to its normal
superconductor counterpart. This amplification is attributed to the strong suppression of local normal Andreev reflection
process (arising from the Cooper pair splitting) from the graphene based superconductor to lead via the same quantum dot, in
comparison to the usual 2D superconductor. Due to the vanishing density of states at the Dirac point of undoped graphene, a doped graphene 
based superconductor is considered here and it is observed that Cooper pair splitting current is very insensitive to 
the doping level in comparison to the usual 2D superconductor. The transport process of non-local spin
entangled electrons also depends on the type of pairing \ie~whether the electron-hole pairing is on-site, inter-sublattice
or the combination of both. The inter-sublattice pairing of graphene causes the maximum non-local Cooper pair splitting current,
whereas presence of both pairing reduces the Cooper pair splitting current.

\end{abstract}
\maketitle 
\section{Introduction}
In recent times, search for spatially separated quantum entangled states (Einstein-Podoloski-Rosen pair~\cite{PhysRev.47.777})
in various condensed matter systems has become an exciting area of research. The generation and detection of entanglement is
the prerequisite for application in the field of quantum computation and information~\cite{nielsen2002quantum}, quantum 
cryptography~\cite{RevModPhys.74.145}, quantum teleportations~\cite{PhysRevLett.70.1895,PhysRevLett.80.1121}, and for testing
Bell's inequality~\cite{RevModPhys.38.447} etc. The two electron bound states inside a superconductor, known as Cooper
pair~\cite{PhysRev.106.162,tinkham1996introduction}, is a natural source of entangled electron pairs. The non-local spin
entangled electrons can be generated out of the superconductor by splitting this Cooper pair by means of Andreev process~\cite{andreev1}.
The latter is an electron-hole conversion phenomena at normal-superconductor interface. After the theoretical proposal of
generating spin-entangled electrons by P. Recher et.al.~\cite{recher2001andreev}, several experiments~\cite{hofstetter2009cooper,PhysRevLett.104.026801,
das2012high,PhysRevLett.115.227003} have been carried out to realize such splitting phenomena. These kind of devices are
generally known as Cooper pair beam splitter (CPS) which consists of two leads attached to the superconductor at two
different points via two different quantum dots in the Coulomb blockade regime. There have been several proposals of detecting
spin entanglement including testing the violation of Bell's inequality~\cite{RevModPhys.38.447,PhysRevLett.91.157002,PhysRevB.66.161320},
shot noise properties~\cite{blanter2000shot,PhysRevB.61.R16303,PhysRevB.70.115330} and Josephson current flowing through
double quantum dots attached to two superconducting leads~\cite{PhysRevB.94.155445} etc.

On the other hand, graphene~\cite{neto2009electronic} is an atomically thin material of carbon atoms and it's low energy
spectrum is described by the massless Dirac equation rather than Schr\"{o}dinger equation. Beenakker, in his seminal paper
~\cite{beenakker2006specular,beenakkerreview} has established that an undoped graphene can exhibit specular Andreev reflection 
(AR) in a normal-superconductor (NS) hybrid junction, which is in contrast to the usual Schr\"{o}dinger type electronic systems 
exhibiting retro type AR. The specular AR in graphene is a direct manifestation of the two band semiconducting nature~\cite{neto2009electronic}
(interband Andreev reflection) with zero band gap. Although, specular AR can also be realized in any low (enough) gap semiconductor 
at low doping~\cite{zhangma}.
Later, J. Cayssol~\cite{cayssol2008crossed} has investigated quantum transport properties in a normal-superconductor-normal 
(NSN) hybrid junction made of graphene monolayer, and predicts that graphene could be a better candidate to realize spin-entangled electrons 
via the Crossed Andreev reflection (CAR) process. Very recently, graphene based CPS has been designed experimentally in order to 
enhance CPS current~\cite{borzenets2016high, PhysRevLett.114.096602} compared to normal superconductor. In the experiment, superconducting 
correlation has been induced in graphene via the proximity effect. They have observed remarkably better performance
of Cooper pair splitting by tuning gate voltage. However, these findings have been naively explained by 
using the theoretical work of P. Recher et al.,~\cite{recher2001andreev} which is based on BCS type 3D normal superconductor with quadratic specturm. 
Hence, a microscopic theoretical analysis of graphene superconductor based CPS geometry is on demand for a better understanding of the 
enhancement of CPS current in it.

In this article, we intend to provide a theoretical analysis of CPS mechanism for a graphene based superconductor
where pairing symmetry is considered to be originated from phonon mediated interaction and tailed by the discussions
regarding proximity induced pairing. We explore the origin behind the enhancement of beam splitting process in
graphene, and discuss the outcome of different kinds of pairing symmetries which are on-site, inter-sublattice
and presence of both. This work is an extension of Ref.~[\onlinecite{recher2001andreev}] to Dirac
supercondutor incorporating the same formalism introduced by them.

We show that the Cooper pair splitting visibility $(\eta)$~\cite{borzenets2016high} is amplified in Dirac like superconductor
(graphene with linear spectrum) in comparison to the usual 2D BCS type superconductor with parabolic dispersion. The visibility 
can be defined as $\eta=I_{CPS}/(I_{CPS}+I_{BG})$~\cite{borzenets2016high} with $I_{CPS}$ being the current via the two different
dots and $I_{BG}$ is the current via the same dot. The types of pairing inside the graphene superconductor play a crucial role in
CPS process. The CPS visibility is minimum when only on-site pairing is present. On the other hand, it becomes maximum when only
inter-sublattice pairing is present. The presence of both types of pairing give rise to an intermediate $\eta$.

The remainder of the paper is organised as follows. The BCS theory for graphene is briefly reviewed in Sec.~\ref{bcs}.
In Sec.~\ref{cps}, we analytically evaluate the Cooper pair splitting current via two different dots as well as 
via the same dot for the graphene CPS geometry. In Sec.~\ref{results}, we discuss the outcome of our analytical results. 
Finally, we summarize and conclude in Sec.~\ref{summary}.

\section{Brief review of BCS Theory of graphene}\label{bcs}
In this section, we briefly review the BCS theory of graphene superconductor as prescribed by B. Uchoa et al., in
Ref.~[\onlinecite{PhysRevLett.98.146801}]. We start with the tight binding Hamiltonian for graphene as
\begin{equation}\label{gra_Ham}
 H_0=-\mu\sum_{i}\hat{n}_{g,i}-t\sum_{\la ij\ra}\sum_{s=\ua\da}(\alpha_{is}^{\dagger}\beta_{js}+H.c) \ .
\end{equation}
Here, $s=\ua,\da$ denotes the spin index, $t$ is the nearest neighbor hoping parameter ($t\simeq 2.6$ eV)
between A and B sublattice, $\alpha_{i}$($\alpha_{i}^{\dagger}$) is the on-site annihilation (creation)
operator of electron in the A sublattice. Similarly for the B sublattice, $\beta_{i}$ ($\beta_{i}^{\dagger}$) is the 
annihilation (creation) oparator. $\mu$ is the chemical potential with $\hat{n_{g,i}}$ being the on-site particle density operator.
After diagonalizing Eq.(\ref{gra_Ham}), the energy dispersion becomes as $\xi_{k}=-t|\gamma_{k}|$,
where $k$ is the 2D momentum and $\gamma_{k}=\sum_{\vec{\delta}}e^{i\vec{k}.\vec{\delta}}$. 
Here, $\vec{\delta}=\{\vec{\delta}_{1},\vec{\delta}_{2},\vec{\delta}_{3}\}$ are the three nearest
neighbor lattice vectors. The low energy approximation at the corner of hexagonal Brillouin zone leads
to the linear Dirac spectrum given by $\xi_{k}=\hbar v_{F}|k|$ with $v_{F}=(3/2)at$-the Fermi velocity.
Now we include electron-hole pairing in graphene via phonon mediated electron-electron interaction,
which can be described by
\begin{eqnarray}\label{int}
 H_{\rm int}&=&\frac{g_0}{2}\sum_{is}[\alpha_{is}^{\dagger}\alpha_{is}\alpha_{i-s}^{\dagger}\alpha_{i-s}+\beta_{is}^{\dagger}
 \beta_{is}\beta_{i-s}^{\dagger}\beta_{i-s}]
 \nonumber\\&+& g_{1}\sum_{\la ij\ra}\sum_{s,s'}\alpha_{is}^{\dagger}\alpha_{is}\beta_{js'}^{\dagger}\beta_{js'} \ ,
\end{eqnarray}
where $g_0$ and $g_{1}$ are the on-site and nearest neighbor electron-electron interaction strength, respectively. 
Introducing two types of superconducting order parameters as (a) $s$-wave: $\Delta=\la \alpha_{i\da}\alpha_{i\ua}\ra=\la \beta_{i\da}\beta_{i\ua}\ra$
(b) $p$-wave: $\Delta_{1,ij}=\la \alpha_{i\da}\beta_{i\ua}-\alpha_{i\ua}\beta_{i\da}\ra$, the interaction terms under mean field approximation
reduces to~\cite{PhysRevLett.98.146801} 
\begin{eqnarray}
 H_{int}&=&E_0+g_0\Delta\sum_{i}[(\alpha_{i\ua}^{\dagger}\alpha_{i\da}^{\dagger}+\beta_{i\ua}^{\dagger}\beta_{i\da}^{\dagger})+H.c]\nonumber\\
 &+&g_{1}\sum_{\la ij\ra}\Delta_{1,ij}[(\alpha_{i\ua}^{\dagger}\beta_{j\da}^{\dagger}-\alpha_{i\da}^{\dagger}\beta_{j\ua}^{\dagger})+H.c]\ .
\end{eqnarray}
Here, $E_0=-g_0\Delta^2-3g_1\Delta_1^2$. In the momentum space, 
$\Delta_{k}=\sum_{ij}\Delta_{1,ij}e^{i\vec{k}.(\vec{r}_{i}-\vec{r}_{j})}=\Delta_{1}\gamma_{k}^{\ast}$.
In close vicinity of Dirac points, nearest neighbour order parameter can be simplified to
$\Delta=(3a/2)\Delta_{1}(k_y+ik_x)$ \ie~with {\bf p+ip} symmetry. In order to decouple the two sublattices,
we employ the following transformations as~\cite{black2007resonating}
\begin{eqnarray}
 \alpha_{ks}&=&\frac{1}{\sqrt{2}}[c_{ks}+d_{ks}]\nonumber\\
 \beta_{ks}&=&\frac{1}{\sqrt{2}}[e^{-i\phi_{k}}(c_{ks}-d_{ks})]
\end{eqnarray}
where $c_{ks}^{\dagger}(d_{ks}^{\dagger})$ creates an electron in the lower (upper) $\pi$-band. Here, $\phi_{k}=\arg(\gamma_{k})$.
Then after the Bogoliubov transformation, BCS Hamiltonian reduces to 
\begin{equation}
H_{BCS}=\sum_{k}E_{(k,\nu=+)}\gamma_{1ks}^{\dagger}\gamma_{1ks}
+E_{(k,\nu=-)}\gamma_{{2ks}}^{\dagger}\gamma_{2ks}
\end{equation}
with Bogoliubov quasiparticle's energy $E_{k,\lambda,\nu}=\lambda E_{k,\nu}$, where $\lambda=\pm$ and
\begin{equation}
 E_{k,\nu}=\sqrt{(\xi_{k}+\nu\mu)^2+(g_0\Delta+\nu g_1\Delta_1|\gamma_{k}|)^2)}.
\end{equation}
Here, $s\equiv\ua\da$ and $\gamma_{1ks}$ ($\gamma_{2ks}$) is the quasiparticle operator, corresponding to 
the energy $E_{k,+}$ ($E_{k,-}$) which can be linked to the annihilation and creation operators as
\begin{eqnarray}
c_{k,\nu,\ua}&=&u_{k,\nu}\gamma_{1k\ua}+v_{\nu,k}\gamma_{1-k\da}^{\dagger}\\
c_{-k,\nu,\da}&=&u_{k,\nu}\gamma_{1-k\da}-v_{\nu,k}\gamma_{1k\ua}^{\dagger} 
\end{eqnarray}
with the quasiparticle weights $u_{k,\nu}=(1/\sqrt{2})(1+\xi_k/E_{k,\nu})^{1/2}$, and $v_{k,\nu}=(1/\sqrt{2})(1-\xi_k/E_{k,\nu})^{1/2}$
. Note that, unlike normal superconductor, graphene superconductor exhibits two kinds of Bogoliubov
quasiparticles in each band with different energies denoted by $\nu$. The appearence of two types of Bogoliubov quasiparticles
with different energies is in complete contrast to usual normal BCS superconductor where Bogoliubov quasiparticle is of one type.
This unusual feature of Bogoliubov quasiparticle in graphene is going to play an important role in graphene based CPS geometry.
Here we investigate three different
cases:\\
(a) Intra-sublattice pairing \ie~$\Delta\neq 0$ and $\Delta_1=0$: In this case, the superconductor is described by the gap
$2|g_0\Delta|$. Note that, this on-site pairing is also equivalent to proximity induced pairing in graphene~\cite{beenakker2006specular}.\\
(b) Inter-sublattice pairing \ie~$\Delta=0$ and $\Delta_1\neq0$:
In this case, the hopping parameter and the chemical potential are renormalized as
\begin{equation}
 t'=\sqrt{t^2+g_1^2\Delta_1^2}
\end{equation}
and 
\begin{equation}
 \mu'=\frac{\mu t}{\sqrt{t^2+g_1^2\Delta_1^2}} \ .
\end{equation}
Hence, the quasiparticle's energy reduces to
\begin{equation}
 E_{k,\nu}=\sqrt{(t'|\gamma_k|+\nu \mu')^2+{\Delta'}^2}
\end{equation}
with gap $\Delta'=\mu g_1\Delta_1/\sqrt{t^2+g_1^2\Delta_1^2}$. Note that, superconducting gap vanishes for undoped graphene
($\mu=0$). So this kind of pairing is intrinsically related to finite doping.\\ 
(c) Presence of both the pairings \ie~ $\Delta\neq 0$ and $\Delta_1\neq 0$: In this case, the energy gap turns out to be
$2|tg_0\Delta-g_1\mu\Delta_1|/\sqrt{t^2+g_1^2\Delta_1^2}$ with renormalized chemical potential
\begin{equation}
 \mu'=(t\mu+g_{0}g_{1}\Delta\Delta_{1})/\sqrt{t^2+g_1^2\Delta_1^2} \ .
\end{equation}
Because of the vanishing density of states at the Dirac point, realization of superconductivity in undoped graphene is difficult.
Hence one has to tune the chemical potential substantially above the Dirac points which can be done by chemically
doping the graphene with metal coating as mentioned in Ref.~[\onlinecite{PhysRevLett.98.146801}].
\begin{figure}[!thpb]
\centering
\includegraphics[height=5cm,width=0.45\textwidth]{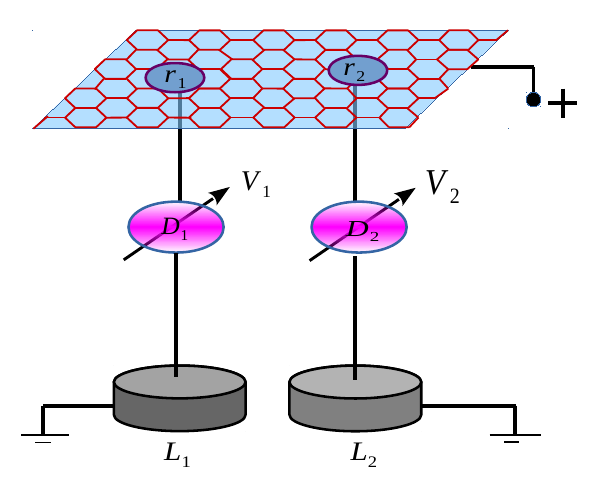}
\caption{(Color online) Schematic of the Cooper pair beam splitter device based on the graphene superconductor.
Two quantum dots are denoted by $D_1$ and $D_2$. Leads are shown by two circular disks and referred by $L_1$ and $L_2$. 
Two gate voltages $V_1$ and $V_2$ are coupled to the two dots to tune the energy levels of the dots.}
\label{beam}
\end{figure}
\section{Graphene based Cooper pair beam splitter}\label{cps}
In this section, we analytically evaluate the Cooper pair splitting current following the route given in Ref.~[\onlinecite{recher2001andreev}]
for the normal 3D BCS superconductor. Before proceeding further we characterize our device with different scattering processes and required conditions. 
The mechanisms which are involved in the beam splitting phenomena are the Andreev processes and Coulomb blockade effects.
We consider that the graphene superconductor is kept at a chemical potential $\mu_S$, and weakly coupled to two separate leads via
two quantum dots as shown schematically in Fig.~\ref{beam}. Note that, to avoid edge effect of graphene, we consider contact points
much inside the graphene sheet. Two quantum dots are denoted by $D_1$ and $D_2$, while the two normal leads are denoted by $L_1$ and
$L_2$, respectively. Two leads are kept at the same chemical potential \ie~$\mu_1=\mu_2=\mu$, where $\mu_1$ and $\mu_2$ refer
to the chemical potential of lead 1 and lead 2, respectively. Transport of entangled  electrons occurs under the bias voltage
$\Delta\mu=\mu_s-\mu_l>0$, where $l=1, 2$ denote lead $L_1$ and $L_2$, respectively. The single particle energy levels $\eps_1$ and $\eps_2$
of the two quantum dots can be tuned externally via gate voltage ($D_1$ and $D_2$) to satisfy resonance condition, described by two particle Breit-Wigner
peak at $\eps_1+\eps_2=2\mu_s$. The latter describes the co-tunneling of two electrons into two different dots. To block the
unwanted correlation between electrons, already present on the quantum dots and electron coming from superconductor, one would
work in the co-tunneling regime in which the number of electrons on the dots are fixed and the resonant levels $\eps_l$ cannot
be occupied. To prevent the spin flip process energy level spacing of quantum dots has to be higher than thermal energy $k_BT$
and bias voltage $\Delta\mu$, where $k_B$ and $T$ are the Boltzmann constant and temperature, respectively. The electron that
enters into the dots from the superconductor must leave the dot to lead much faster than the time scale in which another electron 
arrives into the dots \ie~$|T_{SD}|<|T_{DL}|$, where $T_{SD}$ is the tunneling amplitude from superconductor to dot and $T_{DL}$ is the
tunneling amplitude from dot to superconductor, respectively. The superconducting energy gap also characterizes the time delay between
two successive Andreev tunneling events of the two electrons of a Cooper pair. In order to suppress the single electron tunneling
where the creation of the quasiparticle in the superconductor is a final excited state, one require that $\Delta>\Delta\mu$,$k_{B}T$.
The Hamiltonian of the entire system is described by
\begin{equation}
 H=H_{BCS}+\sum_{I}H_{DI}+\sum_{I}H_{LI}+H_{T}
\end{equation}
with $I=1,2$. Here, the superconductor is described by the graphene BCS Hamiltonian with $\gamma_{\nu}|0\ra_{\nu}=0$. Both dots are
modeled as Anderson-type Hamiltonian given by $H_{DI}=\eps_{I}\sum_{s}d_{ls}^{\dagger}d_{ls}+Un_{\ua}n_{\da}$, where $U$ is the Coulomb
blockade energy. Only resonant levels of the dots
participate in transport phenomenon here.

The leads are assumed to be non-interacting Fermi liquids with the Hmiltonian $H_{LI}=\sum_{ks}\eps_{k}a_{Iks}^{\dagger}a_{Iks}$. 
Tunneling from suerconductor to dots and dots to leads are described by the tunneling Hamiltonian $H_{T}=H_{SD}+H_{DL}$ given as
\begin{equation}
 H_{SD}=\sum_{ls}T_{SD}d_{Is}^{\dagger}\psi_{s}(\vec{r}_{I})+h.c
\end{equation}
and 
\begin{equation}
 H_{DL}=\sum_{lks}T_{DL}a_{Iks}^{\dagger}d_{Iks}+h.c
\end{equation}
The field operators are given by $\psi_{s}(\vec{r}_{I})=\sum_{k}e^{i\vec{k}.\vec{r}}{c}_{ks}$ or $\sum_{k}e^{i\vec{k}.\vec{r}}{d}_{ks}$
depending on type of Bogoliubov quasi particles.

The Cooper pair splitting current from superconductor to the dots is given by
\begin{equation}
 I=2e\sum_{f,i}W_{fi}\rho_i
\end{equation}
with the transition rate
\begin{equation}
 W_{fi}=2\pi|\la f|T(\eps_i)|i\ra|^2\delta(\eps_f-\eps_i).
\end{equation}
Here, $T(\eps_i)=H_{T}G(E)(\eps_i-H_0)$ is the on shell transmission or T-matrix with $G(E)=[\eps_f+i0-H]^{-1}$.
The initial occupation probability for the entire system in states $|i\ra$ is denoted by $\rho_i$.
The T-matrix can be written as a power series in tunnel Hamiltonian as~\cite{recher2001andreev}
\begin{equation}
 T(\eps_i)=H_{T}+H_{T}\sum_{n=1}\big[G(E)H_{T}\big]^n \ .
\end{equation}
The initial state is defined as $|i\ra=|0\ra_{S}|0\ra_{D}|\mu_l\ra$, where $|0\ra$ is the quasiparticle
vacuum for the superconductor. Furthermore, $\gamma_{S}=2\pi\nu_S|T_{SD}|^2$ and $\gamma_{I}=2\pi\nu_I|T_{DL}|^2$
($I=1,2$) denote the tunneling rates between superconductor and dots and between dots and leads, respectively. Also,
$\nu_S$ and $\nu_I$ are the density of states of graphene and the leads, respectively.
\subsection{Current via two dots}
Here, we analytically evaluate the current due to simultaneous transport of two electrons via two different dots. This process is known
as the crossed Andreev reflection in literature. The two electrons coming out of the graphene superconductor can be either
singlet (S=0) or triplet (S=1). The conservation of the total spin S, $[{\bf S^2},H]=0$, guarantees the preservation of the
singlet or triplet states of Cooper pair during the transport process via two dots into the leads. The final states of the two electrons,
in two differet leads, are described by the quantum state $|f\ra=(1/\sqrt{2})[a_{1p\ua}^{\dagger}a_{2q\da}^{\dagger}
\pm a_{1p\da}^{\dagger}a_{2q\ua}^{\dagger}]|i\ra$, where $-$ and $+$ denote singlet and triplet, respectively. Here, $p$
and $q$ are the momentum vector in two leads corresponding to the energy $\eps_p$ and $\eps_q$, respectively. Also,
$a^{\dagger}_{1ps}$ is the creation operator of electon with spin $s$ in lead 1 with momentum $p$, whereas $a^{\dagger}_{2qs}$
denotes the same for lead 2 with momentum $q$. After splitting of the Cooper pair, one electron with spin $\ua(\da)$ migrates
to dot 1 from the contact point $r_1$ of superconductor. The second electron with spin $\da(\ua)$ from contact point
$r_2$ migrates to the dot 2 before the electron with spin $\ua(\da)$ in the dot 1 escapes to lead 1. The matrix element
for the final states being singlet in dots, can be directly obtained following Ref.~[\onlinecite{recher2001andreev}] as
\begin{equation}\label{MSD}
 M_{SD}=\frac{4T_{SD}^2}{\eps_1+\eps_2-i\eta}\sum_{k,\nu}\frac{u_{k,\nu}v_{k,\nu}}{E_{k,\nu}}\cos({\bf k.\delta r})\ ,
\label{msd}
\end{equation}
where ${\bf \delta r=r_1-r_2}$ denotes the separation between the two contact points inside the superconductor from which
electrons 1 and 2 tunnel into the dots. To evaluate the sum over k [see Appendix \ref{appA} for details], we use $u_{k,\nu}v_{k,\nu}=
\Delta/(2E_{k,\nu})$. Note that, the expression in Eq.(\ref{msd}) is similar as Ref.~[\onlinecite{recher2001andreev}] except an additional 
summation over index $\nu$ attributed to two different branches of Bogoliubov quasi particles for graphene superconductor. 
After linearizing the energy dispersion around the Fermi level, the summation can be reduced to

\begin{equation}
  \sum_{k,\nu}\frac{u_{k,\nu}v_{k,\nu}}{E_{k,\nu}}\cos({\bf k.\delta r})=\frac{\pi}{2}\nu^{g}_s\kappa^g(k_F\delta r) \ .
\end{equation}
Here, $k_{F}$ is the Fermi momentum at chemical potential $\mu_s$, and the density of states of graphene
$\nu^{g}_s=\mu_s/[2\pi(\hbar v_F)^2]$. Here, in above equation
\begin{equation}
 \kappa^g(x)=\frac{1}{\pi}[gK_{0}(gx)+g^{\ast}K_{0}(g^{*}x)] \ ,
\end{equation}
where $K_{0}$ is the zeroth order modified Bessel function, $g=[1/(\pi\xi^{g}_lk_F)+i]$ and $\xi^{g}_l=\hbar v_F/(\pi\Delta)$ is the 
superconducting coherence length. Now, if we evaluate the same for usual 2D superconductor (only single energy branch), then

\begin{equation}
 \sum_{k}\frac{u_kv_k}{E_k}\cos({\bf k.\delta r})=\frac{\pi}{2}\nu^{2d}_s\kappa^{2d}(k_F\delta r)\ .
\end{equation}
Here,
\begin{equation}
\kappa^{2d}(x)=\frac{1}{\pi}[K_0(\omega^{\ast} x)-K_0(\omega x)]
\end{equation}
with $\omega=[1+\{2/(\pi k_F\xi^{2d}_l)\}^2]^{1/2}\exp(i\theta/2)$ and $\cot\theta=\pi k_F\xi^{2d}_l/2$. Also,
$\nu^{2d}_s=m^{*}/{2\pi\hbar^2}$ is the density of states of usual 2D electronic systems with $m^*$ is the 
effective mass of electron and $\xi_l^{2d}$ is the coherence length of usual 2D superconductor. 
Note that the coherence length of graphene superconductor is higher than usual 2D
superconductor because of the higher Fermi velocity. It can also be seen that tuanneling amplitude is less sensitive
to the coherence length in usual 2D superconductor than graphene superconductor. An approximate form of the transmission
amplitude for usual 2D superconductor is given in Ref.~[\onlinecite{recher2002superconductor}]. However, to draw
a comparison with graphene we need exact result.

Furthermore, we look into the case of the transmission amplitude from dots to leads. The current from dots to Fermi liquid leads 
is given by~\cite{recher2001andreev}
\begin{equation}
 M_{DL}=-T_{DL}^2\frac{\eps_1+\eps_2-i\eta}{(\eps_1+\eps_q-i\gamma_{1})(\eps_{2}+\eps_p-i\gamma_{2}/2)}.
\end{equation}
However, for our particular case of graphene based superconductor, a factor of $2$ should be multiplied to capture the
contribution from two different kinds of Bogoliubov quasiparticles. After performing integration over momentum $p$
and $q$ of leads, the current via the two dots becomes
\begin{equation}
 I_{CPS}=\frac{e\gamma_s^2\gamma}{(3\mu_s)^2+(\gamma/2)^2}\big|\kappa^{g}(k_F\delta r)\big|^2
\end{equation}
with $\gamma=\gamma_1+\gamma_2$ is the total tunneling rate between dots to leads. Note that, unlike usual 2D
superconductor where density of states is constant and does not depend on energy, it is directly proportional to
the Fermi level in graphene for which a substantial doping or suitable gating is necessary to make Cooper pair
available for transport process, otherwise at the Dirac point $I_{CPS}\simeq 0$ due to the unavailibility of
density of states. For the case of inter-site pairing, we obtain the similar results  with the 
appropriate rescaling of hoping parameter and chemical potential.
\subsection{Current via same dot}
In this subsection, we analytically compute current via the same dots, which can occur via two possible scattering processes, 
(i) one electron with spin-up tunnels to dot 1 from graphene superconductor and then second electron with spin-down also tunnels to dot 1.
When the two electrons with opposite spins are in the same dot, then that process costs an additional energy $U$. This process is the 
usual local Andreev reflection. (ii) One electron tunnels to dot 1 and then go to lead 1 before another electron
from the graphene superconductor arrives at dot 1.
First we consider the second process. Following Ref.~[\onlinecite{recher2001andreev}], we begin with the transmission matrix
\begin{eqnarray}
\la f|T_0|i\ra&=&\sum_{p''s}\la f|H_{DL}|D''s\ra\la Dp''s|\sum_{n=0}\big(\frac{1}{i\eta-H_0}H_{DL}\big)^{2n}|Dp''s\ra\nonumber\\&&
\la Dp''s|\frac{1}{i\eta-H_0}H_{SD}\frac{1}{i\eta-H_0}H_{DL}\frac{1}{i\eta-H_0}H_{SD}|i\ra\ .
\end{eqnarray}
First two matrix elements in the above equation correspond to the transition between dot to lead,
which will be remained same as the leads are kept unchanged, and was already evaluated in 
Ref.~[\onlinecite{recher2001andreev}]. The last matrix element involves the graphene superconductor 
Hamiltonian $(H_{SD})$, for which we evaluate it as [see Appendix \ref{appB} for details]
\begin{eqnarray}
&&\la Dp^{''}s|\frac{1}{i\eta-H_0}H_{SD}\frac{1}{i\eta-H_0}H_{DL}\frac{1}{i\eta-H_0}
\nonumber\\&&H_{SD}|i\ra=\pm\frac{T_{DL}T_{SD}^2\nu^{g}_s}{(\eps_l+\eps_{p''}-i\eta)}\frac{2}{\mu_s}\ .
\end{eqnarray}
Here, +(-) sign corresponds to spin-up (down). Substitution of this expression into the transmission amplitude yields 
\begin{equation}
\la f|T_0|i\ra=-\frac{2\sqrt{2}\nu_s^g(T_{SD}T_{DL})^2(\eps_l-i\gamma_l/2)}{(\eps_l+\eps_p-i\gamma_l/2)
(\eps_l+\eps_{p''}-i\gamma_l/2)}\frac{2}{\mu_s}\ .
\end{equation}
In usual 2D/3D normal BCS superconductor, it gives similar results except $\nu_s^g/\mu_s$ is replaced by $\nu^{2d}_s/\Delta$.
Hence, this process is suppressed by the factor $[2\pi(\hbar v_F)^2]^{-1}$ in graphene based superconductor in comparison to usual 2D
superconductor where it is suppressed by $m^{\ast}(2\pi\hbar^2\Delta)^{-1}$. Due to the higher Fermi velocity in graphene,
the degree of supression of this process is higher in graphene based superconductor than usual 2D BCS superconductor.
\begin{figure}
\includegraphics[width=.48\textwidth,height=6.5cm]{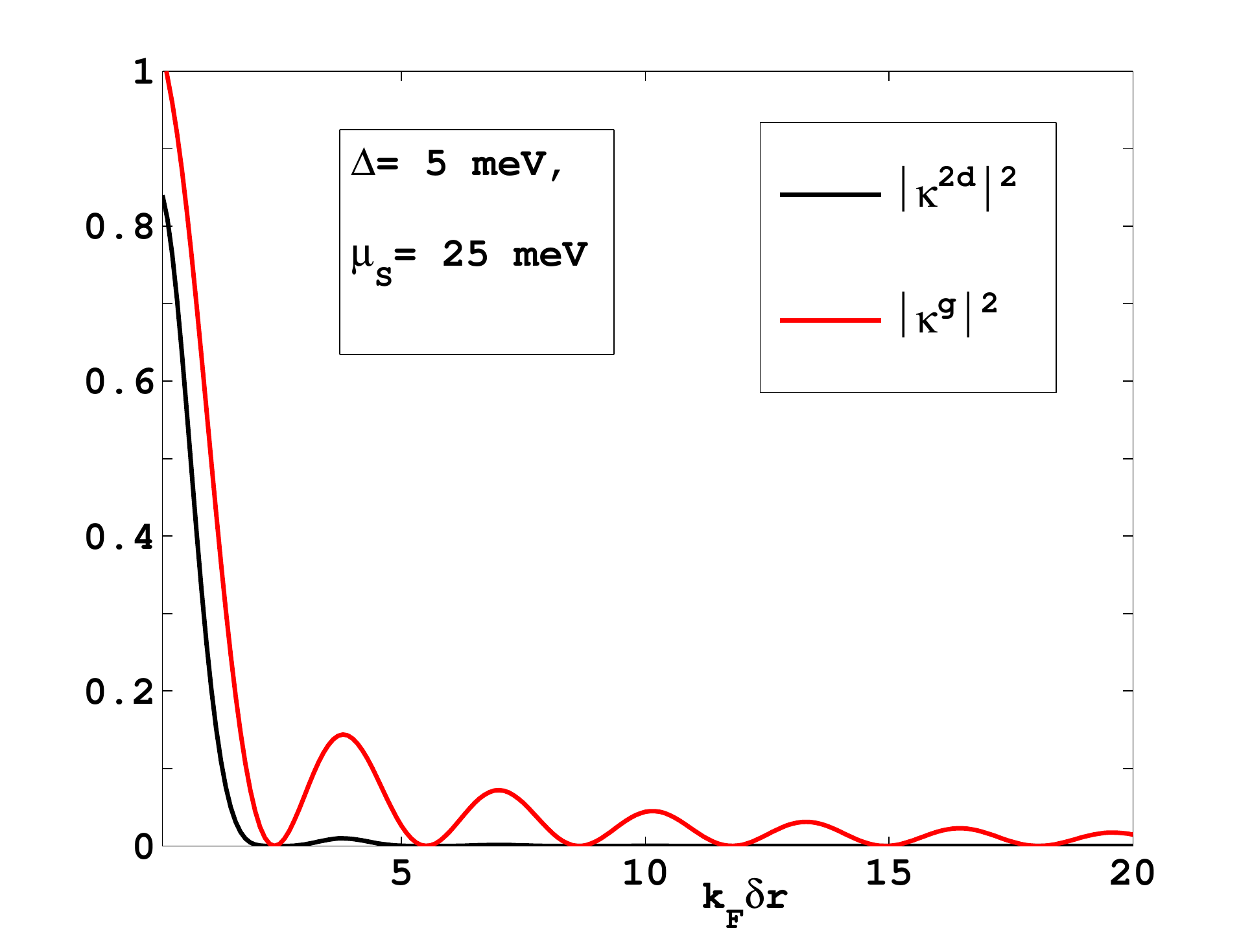}
\caption{(Color online) The behavior of square of the transmission amplitudes for graphene and usual 2D BCS superconductor 
are shown as a function of $k_F\delta r$. Here, $\delta r$ is the seperation between the two electrons of the Cooper pair inside 
the graphene sheet.}
\label{noise_plot}
\end{figure}

\begin{figure*}
\centering
\subfigure{\includegraphics[width=.48\textwidth,height=6.25cm]{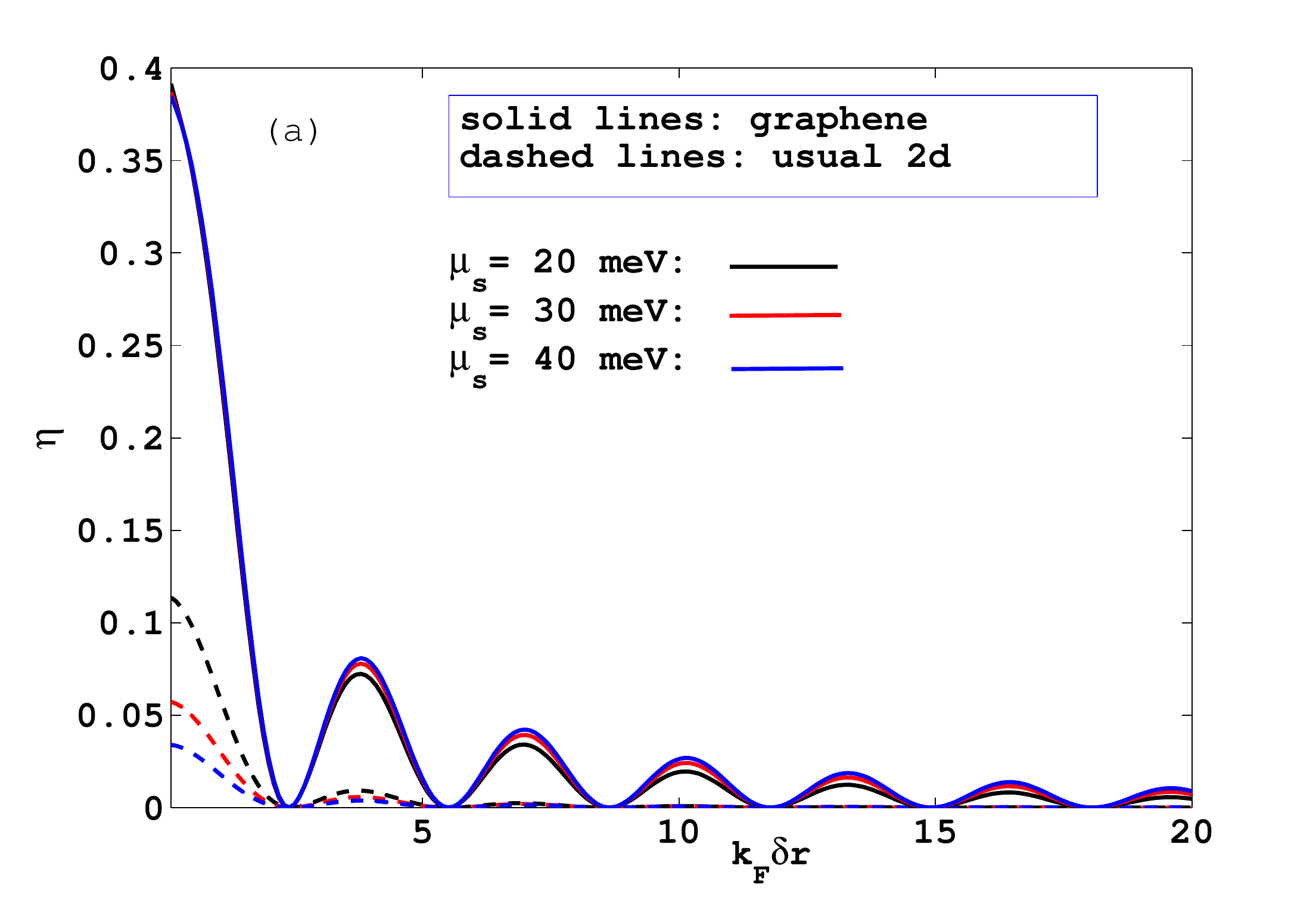}}
\subfigure{\includegraphics[width=0.48\textwidth,height=6.25cm]{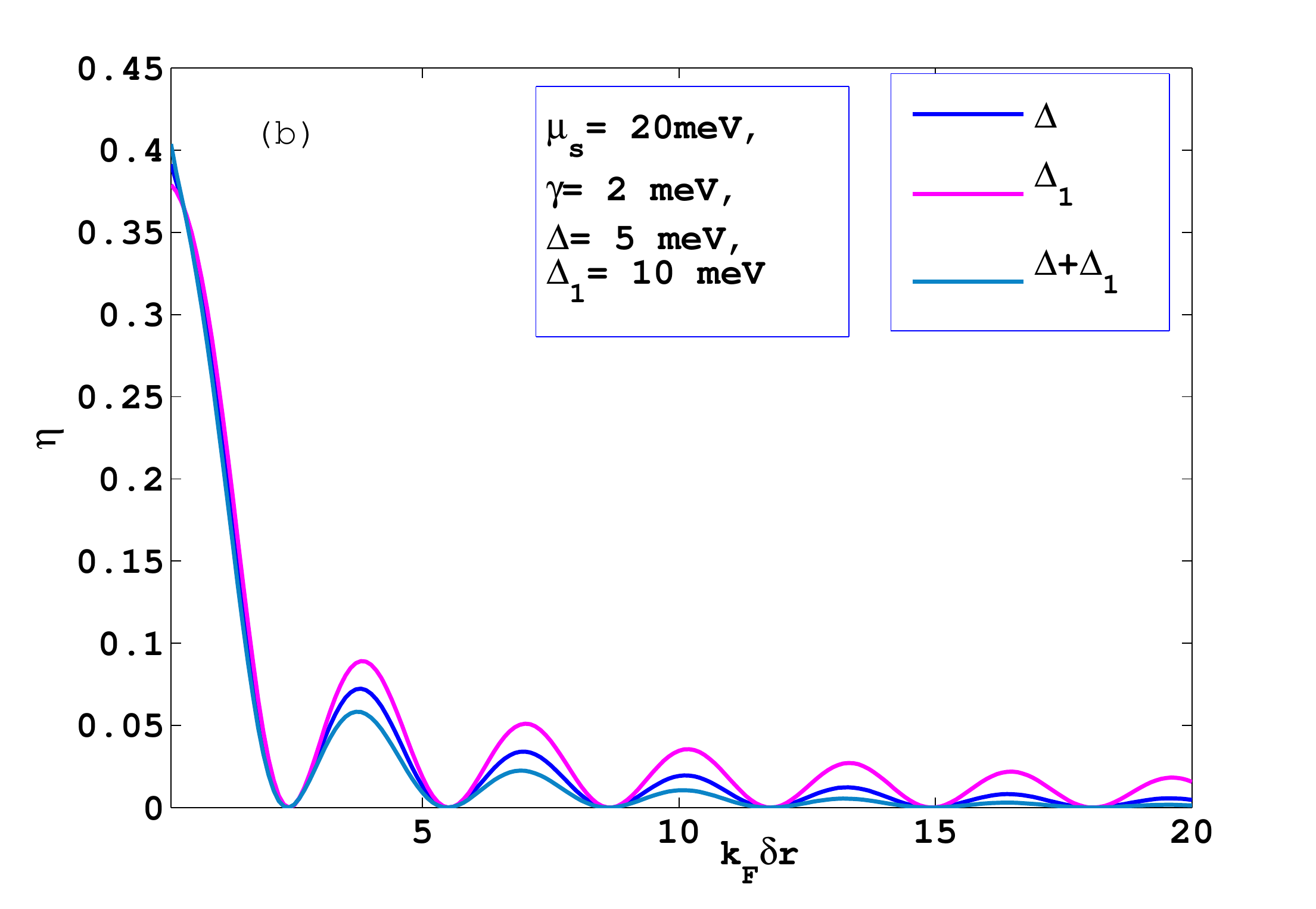}}
\caption{(Color online) CPS visibility ($\eta$) is illustrated as a function of $k_F\delta r$. In panel (a) results are shown for graphene and usual 2D
 superconductor together for various values of doping. On the other hand, in panel (b) results are demonstrated for different kinds of pairing symmetry 
in graphene based superconductor.}
 \label{CPS}
\end{figure*}
Now, we consider the first case, where two electrons tunnel together from the superconductor to dot and because of the Coulomb
blockade phenomena it costs additional energy $U$. In this process, the matrix element involving superconducting Hamiltonian can be evaluated
by just replacing $\Delta$ by $U/\pi$ in normal superconductor. However, in graphene we evaluate it to be as
\begin{eqnarray}
 &&\la Dp^{''}s|\frac{1}{i\eta-H_0}H_{DL}\frac{1}{i\eta-H_0}H_{SD}\frac{1}{i\eta-H_0}
\nonumber\\&&H_{DL}|i\ra=\nu^{g}_{s}\frac{\Delta}{2U\mu_s}\sum_{\nu}\ln\big[1+\big(\frac{\eps_c-\nu\mu_s}{\Delta}\big)^2\big]
\end{eqnarray}
Here, $\eps_c$ is the energy cut-off inside the superconductor. Due to the cummulative effect of these two processes,
the current via the same dot is found to be 
\begin{eqnarray}
 I_{BG}=2e\frac{\gamma_S^2\gamma}{A}B\ ,
\end{eqnarray}
where
\begin{equation}
 B=\frac{2}{\pi\mu_s}+\frac{\Delta}{2U\pi\mu_s}\sum_{\nu}\ln\big[1+\big(\frac{\eps_c-\nu\mu_s}{\Delta}\big)^2\big]\ .
\end{equation}
So the efficiency of Cooper pair splitting for graphene based superconductor (using resonance condition $\eps_1+\eps_2=2\mu_{s}$) becomes
\begin{equation}
 \frac{I_{CPS}}{I_{BG}}=\frac{A}{2B}|\kappa^{g}(k_F\delta r)|^2.
\end{equation}
with 
\begin{equation}
 A=\frac{(3\mu_s)^2+(\gamma/2)^2}{[6\mu_S^2+(\gamma/2)^2]^2+[5\mu_s\gamma/2]^2}
\end{equation}
and for usual 2D superconductor, the same turns out to be
\begin{equation}
 \frac{I_{CPS}}{I_{BG}}=\frac{A\mathcal{E}^2}{2}\big|\kappa^{2d}(k_F\delta r)\big|^2
 \end{equation}
with $1/\mathcal{E}=1/U+1/(\pi\Delta)$.
\section{Results and discussion}\label{results}
In this section, we discuss our results for graphene based Cooper pair beam splitter geometry in comparison to that of the
usual 2D BCS superconductor. First, we show a comparative behavior of tunneling probability for the process A (two electrons tunnel
via two different dots) in Fig.~\ref{noise_plot}. Note that, we use $\Delta$ and $\Delta_1$ in units of $g_0$ and $g_1$, respectively
for numerical plots. Both of them exhibits oscillatory behavior. Nevertheless, in case of normal 2D superconductor,
the amplitude of oscillation decays much faster than graphene supercondutor. It also shows that the tunneling probability for process A
in graphene is higher in magnitude compared to usual 2D superconductor. The Cooper pair splitting probability via two dots survives for 
a relatively wide range of the distance ($\delta r$) between two dots-superconductor contact points as far as the graphene is concerned. 
The origin of this survival can be attributed to the large coherence length and the presence of two types of Bogoliubov
quasiparticles in graphene instead of one type of them in normal 2D superconductor. Since the non-local transport occurs via Crossed
Andreev reflection process, as pointed out in Sec.~\ref{cps}, to draw a clear comparison of performances
between graphene based CPS and usual 2D superconductor based CPS
we explore the CPS visibility, introduced in Ref.~[\onlinecite{borzenets2016high}], as
\begin{equation}
 \eta=\frac{I_{CPS}}{I_{CPS}+I_{BG}} \ .
\end{equation}

In the second case of process B (two electrons tunnel sequencially via same dot), we find that non-local transport is suppressed
by $\nu_{s}^{g}/\mu_s=[2\pi(\hbar v_F)^2]^{-1}$, whereas in normal 2D BCS superconductor it is $\nu^{2d}_s/\Delta=m^{\ast}
/[2\pi\hbar^2\Delta]$. Such suppression in graphene is governed by the factor $1/v_F^2$ and in normal 2D supercoductor it's counterpart 
is $m^{\ast}/\Delta$. Hence, it can be clearly understood that suppression in graphene is much stronger than usual 2D superconductor as
$v_F>\sqrt{\Delta/m^{\ast}}$. On the other hand, the first case of process B, where two electrons tunnel via same dot
simultaneously, costs additional energy $U$ and is suppressed by $\nu^{2d}_s/U$ in usual 2D superconductor.
On the contrary, in case of graphene based superconductor it is $\nu_s^g\frac{\Delta}{2U\mu_s}\sum_\nu\ln\big[1
+\big(\frac{\eps_c-\nu\mu_s}{\Delta}\big)^2\big]$. Note that, in the latter case, suppression does not differ
significantly in two systems for the same density of states and $\eps_c\sim30\Delta$.

In Fig.~\ref{CPS}, we show CPS visibility as a function of $k_F\delta r$. A comparative analysis between the graphene
and the usual 2D superconductor based CPS is demonstrated in Fig.~\ref{CPS}(a). It can be observed that CPS visibility in graphene
superconductor is higher in magnitude and insensitive to the change of chemical potential than normal 2D superconductor
case. The origin of higher CPS visibility can be understood from the fact that current via the same dot is more
suppressed in graphene compared to the usual 2D superconductor as explained earlier. In addition to that, higher
transmission probability via two different dots in graphene based superconductor is also responsible for higher CPS
visibility than normal 2D BCS superconductor. In Fig.~\ref{CPS}(b), we also illustrate the features of CPS visibility in graphene
superconductor with $k_F\delta r$ but for different kinds of pairing symmetry. We consider inter-sublattice
pairing $(\Delta_{1})$ is stronger than on-site pairing $\Delta$ as in isotropic $k$-space on-site pairing is less
favoured due to strong Coulomb repulsion~\cite{jpcm_annica}. It is observed that maximum CPS visibility is acheived
only when inter-sublattice pairing is present. The on-site pairing gives rise to relatively less CPS visibility in comparison
to inter-sublattice pairing. In another situation, when both types of pairing are present, CPS visibility exhibits a minimum value in
comparison to other two individual pairings. Note that, so far our discussion is restricted to phonon-mediated pairing. The variation
of CPS visibility with respect to pairing symmetry is related to the effective superconducting gap as well as effective
chemical potential as discussed in the last paragraph of Sec. \ref{bcs}. The proximity induced pairing can be captured
by the on-site pairing, which has been considered in Ref.~[\onlinecite{beenakker2006specular}] in analyzing specular
Andreev reflection in graphene. The inter-sublattice pairing arises exclusively due to phonon-mediated superconductivity
in graphene. On the other hand, intra-sublattice pairing can be present in both mechanisms (proximity induced
superconductivity as well as phonon mediated electron-electron interaction induced superconductivity).

Here we present a comparative discussion between our theoretical results and the experiments (Refs.~[\onlinecite{borzenets2016high, PhysRevLett.114.096602}]). 
First, in experimental set up (Refs.~[\onlinecite{borzenets2016high, PhysRevLett.114.096602}]), superconductivity has been induced in a graphene sample via the proximity effect. 
They have used $\rm Al$ (normal BCS superconductor) to induce superconductivity in graphene monolayer. On the other hand, we have considered various microscopic phonon 
mediated pairing symmetries in graphene. Nevertheless, the proximity induced gap is analogous to our on-site (intra-sublattice) pairing gap $\Delta$,
for which our analysis can also be mapped to proximitized graphene based CPS geometry. Secondly, in Ref.~[26], there is evidence of having 
visibility ($\eta$) ranging from $0.5-0.86$ which is quite justified by our analysis also considering low $k_{F}$ and intra-sublattice
pairing gap $\Delta$. However, in those experiments (Refs. [\onlinecite{borzenets2016high, PhysRevLett.114.096602}]), two quantum dots are also fabricated on 
the same sample \ie~superconducting graphene lead and dots are in the same plane. In those experimental CPS devices, edge effects cannot be ignored.
On the other hand, in order to avoid the disturbances caused by the edge effects, we have modeled our CPS device in such way that
the two quantum dots are directly tunnel coupled to the bulk of the graphene superconductor, as depicted in Fig.~\ref{beam}.

Finally we discuss if graphene based superconductor can have any significant impact on electrical noise measurement
as a test of entanglement, proposed by P. Samuelson et al., in Ref.~[\onlinecite{PhysRevB.70.115330}]. It has already been 
pointed out in Ref.~[\onlinecite{PhysRevB.70.115330}], that the form of tunneling amplitude from the supercondutor to leads
does not play any significant role in noise measurement. The auto-correlation between two electrons in 
normal Andreev process appears to be~\cite{PhysRevB.70.115330}
\begin{equation}\label{auto}
 S_{AA}=S_{BB}=\frac{4e^2}{h}\int_{-eV}^{eV}dE[\{1+2RT\}|A(E)|^2+2RTA(E)A^{\ast}(-E)].
\end{equation}
On the other hand, the cross-correlation between two entangled electrons via the CAR process can be written as
\begin{equation}\label{cross}
S_{AB}=S_{BA}=\frac{4e^2}{h}\int_{-eV}^{eV}dE[\{T^2+R^2\}|A(E)|^2-2RTA(E)A^{\ast}(-E)]
\end{equation}
with
\begin{equation}
A(E)=\frac{|\kappa^{g/2d}|\gamma}{(E+\eps_1-i\gamma/2)(-E+\eps_2-i\gamma/2)}\ ,
\end{equation}
where, $R=|r|^2$ and $T=1-R$. Note that, the above expression (Eq.(\ref{cross})) is valid for two electrons tunneling through two different dots.
Also, here $E=\eps_q=\eps_p$. In the above Eqs.~(\ref{auto}-\ref{cross}), the energy 
integration of the last term yields 
\begin{equation}
 \int dE A(E)A^{\ast}(-E)=\frac{4\pi|\kappa^{g/2d}|^2\gamma}{4\mu_S^2+\gamma^2}\ .
\end{equation}
This integration determines the degree of bunching which corresponds to the suppression of cross-correlations and 
enhancement of auto-correlations. The bunching behavior of shot noise is the indication of spin singlet electrons coming out of the 
entangler, which can be detectable in experiment. The degree of bunching in the shot noise for graphene and usual 2D superconductor
is goverened by the strength of $\kappa^g$ and $\kappa^{2d}$, respectively. From the Fig.~\ref{noise_plot},
it can be seen that tunneling amplitude for graphene is relatively stronger in comparison to usual 2D superconductor.
Hence the degree of bunching in noise measurement in graphene based beam splitter would be much stronger in comparison
to usual 2D superconductor, making the possible entangled state detection more feasible in case of graphene.
\section{Summary and Conclusions}\label{summary}
To summarize, in this article, we investigate the CPS device based on a graphene superconductor. Our analysis is motivated
by two recent experiments on graphene based Cooper pair splitter device~\cite{borzenets2016high,PhysRevLett.114.096602}.
We use the fact that unlike normal BCS type superconductor, the hexagonal lattice structure of graphene exhibits two types of 
Bogoliuobov quasipartiles with different energy. We find that graphene based superconductor can amplify the Cooper pair
beam splitting visibility, which is also the main claim of the experiments. The amount of amplification also depends on
the type of pairing. The inter sublattice pairing, originated from the electron-phonon interaction in graphene, causes maximum 
beam splitting visibility in comparison to on-site pairing. The latter type of pairing can arise due to both proximity induced superconductivity
and electron-phonon interaction. However, for experimental situation, the proximity induced superconductivity is the only realistic possibility~\cite{borzenets2016high,PhysRevLett.114.096602}.
When both types of pairing are considered, CPS visibility exhibits a minimum. We also observe that the origin behind this amplification 
lies in the fact of strong suppression of electron tunneling via the same dots and corresponding enhancement of CPS current via two different dots 
in comparison to the normal 2D BCS superconductor. 
We also notice that CPS visibility is very insensitive to doping level in graphene in comparison the normal 2D superconductor.
Finally, we discuss that the degree of electrical noise bunching, a signature of entagled states,
is expected to be stronger for graphene based CPS device rather than usual 2D superconductor.
\begin{acknowledgements}
SFI acknowledges Arijit Kundu for useful discussions.
\end{acknowledgements}

\begin{appendix}
\section{Tansmisson matrix for electrons tunneling from superconductor to two different dots}
\label{appA}
Here, we simplify the matrix element corresponding to the CPS current via two different dots (see Eq.(\ref{MSD})) as
\begin{equation}
 M_{SD}=\frac{4T_{SD}^2}{\eps_1+\eps_2-i\eta}\sum_{k,\nu}\frac{u_{k,\nu}v_{k,\nu}}{E_{k,\nu}}\cos({\bf k.\delta r}) \ ,
\end{equation}
The summation over $k$ can be evaluated as follows:
\begin{eqnarray}
\sum_{k,\nu}\frac{u_{k,\nu}v_{k,\nu}}{E_{k,\nu}}\cos({\bf k.\delta r})=\frac{\Delta}{4\pi}
\sum_{\nu}\int_0^{\infty}\frac{kdk}{E_{k,\nu}^2}J_{0}(k\delta r) \ .
\label{A1}
\end{eqnarray}
Here, $J_0(k\delta r)$ is the zeroth-order Bessel function of first kind. We have also
used $u_{k,\nu}v_{k,\nu}=\Delta/2E_{k,\nu}$. Note that $\Delta$ is the superconducting gap, and for different types
of pairing it has to be rescaled accordingly as discussed in Sec.~\ref{bcs}. Here, we use the well known convolution
theory for Laplace transformation 
\begin{equation}
 \int_0^{\infty}f(x)h(x)dx=\int_0^{\infty}(\mathcal{L}f)(s).(\mathcal{L}^{-1}h)(s)ds \ ,
\end{equation}
to evaluate the integral in Eq.(\ref{A1}). After using $\mathcal{L} J_0(k\delta r)=1/\sqrt{s^2+\delta r^2}$ and 
the identity of modified Bessel function of zeroth order
\begin{equation}
K_{0}(g x)=\int_0^{\infty}\frac{\cos(g\rho)}{\sqrt{\rho^2+x^2}}d\rho \ ,
\end{equation}
it is straighforward to have
\begin{eqnarray}
  \sum_{k}\frac{u_{k,\nu}v_{k,\nu}}{E_{k,\nu}}\cos({\bf k.\delta r})=\frac{\pi}{2}
  \nu^{g}_s\kappa^{g}(k_F\delta r) \ .
\end{eqnarray}
Similarly for usual 2D superconductor
\begin{eqnarray}
\sum_{k}\frac{u_{k}v_{k}}{E_{k}}\cos({\bf k.\delta r})=\frac{\Delta}{4\pi}
\sum_{}\int_0^{\infty}\frac{kdk}{E_{k}^2}J_{0}(k\delta r) \ .
\end{eqnarray}
Following the same approach, we obtain
\begin{equation}
 \sum_{k}\frac{u_kv_k}{E_k}\cos({\bf k.\delta r})=\frac{\pi}{2}\nu^{2d}_s\kappa^{2d}(k_F\delta r) \ .
\end{equation}

 \section{Evaluation of the transmission matrix for two electrons tunneling via the same dot}
 \label{appB}
 To evaluate the following matrix elements
\begin{equation}
\la Dp^{''}|\frac{1}{i\eta-H_0}H_{SD}\frac{1}{i\eta-H_0}H_{DL}\frac{1}{i\eta-H_0}H_{SD}|i\ra
\end{equation}
we use the following two complete sets of vector 
\begin{equation}
\sum_{k''p' s}\gamma_{k''s}^{\dagger}a_{p'-s}^{\dagger}|i\ra\la i|a_{p'-s}\gamma_{k''s}=1
\end{equation}
\begin{equation}
\sum_{k's}\gamma_{k's}^{\dagger}d_{-s}^{\dagger}|i\ra\la i|d_{-s}\gamma_{k's}=1
\end{equation}
between $H_{SD}$ and $H_{DL}$; and between $H_{DL}$ and $H_{SD}$, respectively.
Then we obtain
\begin{eqnarray}
&&\la Dp^{''}s|\frac{1}{i\eta-H_0}H_{SD}\frac{1}{i\eta-H_0}H_{DL}\frac{1}{i\eta-H_0}H_{SD}|i\ra\nonumber\\&&=
\pm\frac{T_{DL}T_{SD}^2}{(\eps_l+\eps_{p''}-i\eta)}\sum_{k,\nu}\frac{1}{E^{3}_{k,\nu}}\ ,
\end{eqnarray}
where $\sum_{k,\nu}[E_{k,\nu}]^{-3}=2\nu^{g}_s/\mu_s$. In case of normal 2D superconductor,
$\sum_{k}[E_{k}]^{-3}=\nu^{2d}_s/\Delta$.
In the other case, when two electrons migrate together to the same dot from superconductor, 
then this process costs an additional Coulomb energy $U$. Hence, the corresponding matrix element yields
\begin{eqnarray}
  &&\la Dp^{''}s|\frac{1}{i\eta-H_0}H_{DL}\frac{1}{i\eta-H_0}H_{SD}\frac{1}{i\eta-H_0}H_{DL}|i\ra\nonumber\\
  &=&\nu_{s}^g\frac{\Delta}{2U\mu_s}\sum_{\nu}\big[\ln\big\{1+\big(\frac{\eps_c+\nu\mu_s}{\Delta}\big)^2\big\}\big] \ .
\end{eqnarray}

\end{appendix}
\bibliography{bibfile}{}

\end{document}